\documentclass{sig-alternate}

\usepackage{wrapfig}
\usepackage{balance}  
\usepackage[lined]{algorithm2e}
\usepackage{multirow}
\usepackage{subfigure}
\usepackage{graphicx}
\usepackage{caption}
\usepackage{amsmath}
\usepackage{amssymb}
\usepackage{amsfonts}
\usepackage{xspace}
\usepackage{url}
\usepackage{color}
\usepackage{hyperref}

\begin{document}
\conferenceinfo{KDD '15}{Sydney, Australia.  Aug 10 -- 13, 2015}

\title{Facts and Fabrications about Ebola: A Twitter Based Study}

\numberofauthors{5}
\author{
\alignauthor
Janani Kalyanam\\
       \affaddr{Dept. of ECE}\\
       \affaddr{Univ. of California, San Diego}\\
       \email{jkalyana@ucsd.edu}
\alignauthor
Sumithra Velupillai \\
       \affaddr{Dept. of Computer and Systems Sciences}\\
       \affaddr{Stockholm University, Sweden}\\
       \email{sumithra@dsv.su.se}
\and
\alignauthor Son Doan \\
       \affaddr{Dept. of Biomedical Informatics}\\
       \affaddr{Univ. of California, San Diego}\\
       \email{sodoan@ucsd.edu}
\alignauthor Mike Conway\\
       \affaddr{Dept. of Biomedical Informatics}\\
       \affaddr{Univ. of Utah, Salt Lake City}\\
       \email{mike.conway@utah.edu}
\alignauthor Gert Lanckriet\\
       \affaddr{Dept. of ECE}\\
       \affaddr{Univ. of California, San Diego}\\
       \email{gert@ece.ucsd.edu}
}

\maketitle
\begin{abstract}
Microblogging websites like Twitter have been shown to be 
immensely useful for spreading information on a global scale within seconds. 
The detrimental effect, however, of such platforms is that misinformation 
and rumors are also as likely to spread on the network as credible, 
verified information \cite{Castillo:2011}. From a public health standpoint, the spread 
of misinformation creates unnecessary panic for the public. We recently 
witnessed several such scenarios during the outbreak of Ebola in 2014 \cite{oyeyemi2014ebola,time:2014}. 
In order to effectively counter medical misinformation in a timely manner, our goal here 
is to study the nature of such misinformation and rumors in the United States
during fall 2014 when a handful of Ebola cases were confirmed in North
America.

It is a well known convention on Twitter to use
hashtags to give context to a Twitter message (a tweet). 
In this study, we collected approximately 47M tweets from the 
Twitter streaming API related to Ebola. Based on hashtags, we propose a method to 
classify the tweets into two sets: \emph{credible} and \emph{speculative}. 
We analyze these two sets and study how they differ in terms of 
a number of features extracted from the Twitter API.
In conclusion, we infer several interesting differences between
the two sets.  We outline further potential directions to 
using this material for monitoring and separating speculative 
tweets from credible ones, to enable improved public health information.
\end{abstract}

\category{J.3}{Life and Medical Sciences}{Miscellaneous}

\terms{experimentation}

\keywords{Twitter, Ebola, Public health, Misinformation, Rumors, Digital epidemiology}

\section{Introduction}
\label{sec:introduction}
Following the first verified case of Ebola virus in the 
United States in the fall of 2014, there was an explosion of
messages related to the virus on Twitter. Even the slightest suspicion of a 
potential case lead to false rumors and misinformation \cite{time:2014}.
In fact, \cite{oyeyemi2014ebola} found that the majority of the tweets about Ebola
from Liberia, Nigeria and Guinea contained misleading information.
Once disseminated, such misleading information can spread like
wildfire, and create panic amongst the public.
A recent survey-based research found as well that those
with the most knowledge about Ebola gleaned this knowledge from
the internet \cite{rolison2015knowledge}.  It will therefore be
in the primary interest of public health organizations
(e.g., Center for Disease Control)
to correct any misleading information and false rumors on the web
as quickly as possible.

Our aim in this study is to understand what sparks misinformation, what its 
characteristics are, and how it spreads in social media.  The hope
here is that, a deeper
understanding of such concepts will help identify rumors, localize their source
and hence combat rumor diffusion.  
As a first step, we aim to understand the characteristics of
Ebola related misinformation as it is manifested on the microblogging service Twitter.
We describe a dataset that we collected from the Twitter 
Streaming API in early October of 2014.  We describe the 
characteristics of the dataset and explore some preliminary aspects 
of what constitutes truth and rumor in this setting.

\section{Related Studies}
\label{sec:related_studies}
The social networking world (like Twitter, Facebook e.t.c.)
is an open forum to post and share information, news, personal thoughts and experiences.
For data researchers, such services
have opened up unforeseen possibilities because of access to the data posted
by different kinds of users \cite{doan2012analysis}.
Researchers have found
such data immensely useful to better understand several health issues.
For example, \cite{DeChoudhury:2014} study postpartum depression from Facebook data, 
\cite{myslin2013using} study the effects of newer tobacco products from 
Twitter feeds.
Such research that focuses on the data derived from the internet (typically from
of web searches or social networking forums) is called Digital Disease Surveillance \cite{hartley2010landscape,anema2014digital}, also known as Infodemiology
\cite{eysenbach2009infodemiology,eysenbach2006infodemiology, doan2012enhancing}
or Digital Epidemiology \cite{xie2014correlation}.

However, since the data used for research in such areas hails from
open forums, it tends to be extremely noisy, both in terms of
the credibility and in terms of relevance.
For such reasons, the study of information credibility on social media has 
garnered significant interest in recent years.  \cite{Castillo:2011} performed a feature 
based study of information credibility on Twitter.  
They collected several news events, and obtained crowdsourced ground 
truths for which news items were categorized as credible or rumors, 
and employed supervised classification strategies to differentiate 
between credible and rumor tweets.  \cite{Gupta:WWW_2013} performed a similar study to identify fake 
images during hurricane Sandy in 2012.  

However, in contrast to the above studies, our focus is on public health related
rumors, particularly in the context of concerns in the United States regarding
the spread of Ebola from West Africa in fall 2014
\footnote{We note that the work of \cite{oyeyemi2014ebola}, while similar in spirit, 
is very different in terms of approach and findings.  We focus more on studying rumor in Ebola
from Twitter based features.}.  Additionally, in the future, we also hope to understand 
how the propagation of rumor differs from the propagation of 
credible information in public health related issues.

\section{Data Description}
\label{sec:data_description}
\begin{figure}
\begin{center}
    \centering
    \includegraphics[width=0.5\textwidth]{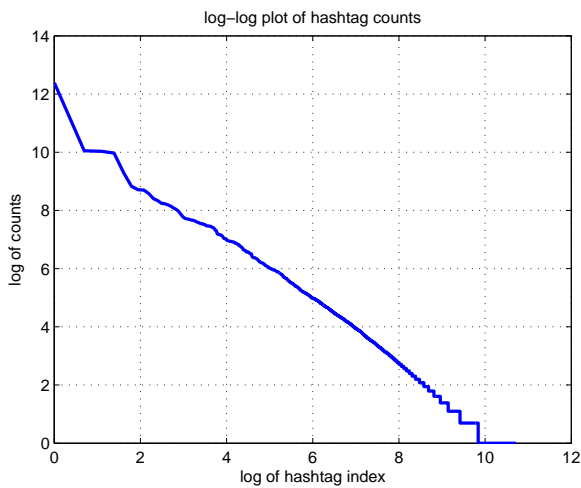}
\end{center}
  \caption{\textbf{\small{This figure depicts the hashtags-vs-counts in a log-log scale.  The hashtags were
sorted in descending order according to their counts.  The x-axis is the log of the index of the
sorted list.  The y-axis is the log of the actual counts.  It can be observed that in a log-log
plot, the decay of the hashtag counts is linear.}}}
\label{fig:hashtag_counts}
\end{figure}
\begin{table}
\centering
\begin{tabular} {|c|c|}
\hline
\textbf{hashtag} & \textbf{count} \\
\hline
\texttt{\#ebola} & 1610343 \\
\texttt{\#news} & 106070 \\
\texttt{\#tcot} & 74413 \\
\texttt{\#breaking} & 41416 \\
\texttt{\#health} &  39500\\
\texttt{\#cdc} &  38172\\
\texttt{\#ebolaoutbreak} &  34656\\
\texttt{\#salvemosaexcalibur} &  28278\\
\texttt{\#obama} &  26840\\
\texttt{\#ebolaresponse} & 26505 \\
\texttt{\#anamatoimision} &26266 \\
\texttt{\#africa} & 22349 \\
\texttt{\#usa} &  20870\\
\texttt{\#isis} & 20418 \\
\texttt{\#dallas} & 17292 \\
\texttt{\#liberia} & 15693 \\
\texttt{\#nigeria} & 14839 \\
\texttt{\#texas} & 13892 \\
\texttt{\#nyc} & 13034 \\
\texttt{\#factsnotfeat} &13021 \\
\hline
\end{tabular}
\caption{\textbf{\small{This table lists the top-20 hashtags in the dataset sorted in descending order according to their counts.}}}
\label{tab:hashtag_counts}
\end{table}

\begin{table}
\centering
\begin{tabular} {|c|c|}
\hline
\textbf{speculative} & \textbf{credible} \\
\hline
\texttt{\#falsenews} & \texttt{\#cdc} \\
\texttt{\#ebolafootballchants} & \texttt{\#breakingnews} \\
\texttt{\#cdcwhistleblower} & \texttt{\#cnn} \\
\texttt{\#thewalkingdead} & \texttt{\#foxnews} \\
\texttt{\#gossip} & \texttt{\#ebolaresponse} \\
\texttt{\#the\_walking\_dead} & \\
\texttt{\#ebolazombie} &  \\
\texttt{\#walkingdead} &  \\
\texttt{\#betterebolaczars} &  \\
\texttt{\#andnowihaveebola} &  \\
\texttt{\#scarystoriesin5words} &  \\
\texttt{\#zombie} &  \\
\texttt{\#twitterjoke} &  \\
\texttt{\#ebolajokes} &  \\
\texttt{\#ebolahoax} &  \\
\texttt{\#lies} &  \\
\hline
\end{tabular}
\caption{\textbf{\small{This table lists the credible and the speculative hashtags used in the study.}}}
\label{tab:credible_speculative}
\end{table}

In the month of October 2014, using the Twitter streaming API, 
we downloaded approximately 47M tweets which had any of the following
keywords: \emph{`ebola',`ebov',`ebolavirus', `sudan virus',`reston virus',`bundibugyo virus'}.  
As a first step in understanding the underlying differences 
between messages that are thought to be rumor and those 
that are thought to be credible, we set off to produce 
two sets of Twitter messages (tweets); one set of which can be thought
of as a credible or confirmed set of tweets,
and the other which can be thought of as more non-serious
or speculative in nature.  We will hereby refer to
the two sets as \emph{credible} and \emph{speculative}.
A surest way of knowing which tweets were credible and which 
tweets were rumorous is to have humans annotate each and every tweet.  
This was in fact the technique adopted by \cite{Castillo:2011}.  
However, as a preliminary experiment, we adopt to a less labor
intensive method of obtaining these two sets.  

On Twitter (and on many other social networking services), 
hashtags are a sequence of non-whitespace characters which follow the `\#' 
sign.  It is a popular convention on Twitter to embed a hashtag 
in a tweet to identify what the tweet is about.  
In some sense, these hashtags can be thought of as annotations given 
by users to give context to the tweet.  In this study, we use these 
hashtags as ground truth for what is correct information, and what 
is possibly misinformation about Ebola.  In the area of topic modelling, 
there have been works that have employed this idea to obtain ground truths \cite{Tsur:2012}.

In our dataset, out of the 47M tweets, approximately 2.7M tweets 
contained a hashtag embedded in them.  Although there were approximately 
184.8K unique hashtags in the dataset, only very few of them were used frequently.  
Out of the 184.8K unique hashtags, only about 400 hashtags had occurred in more than 
1000 tweets.  Moreover, about 110K of the hashtags were 
used only in one tweet.  In other words, most hashtags were used very 
sparsely in tweets.  The most frequent hashtag was \texttt{\#ebola}, and 
had occurred in approximately 1.6M times.  Hence, about 60\% 
of the tweets with hashtags indeed contained \texttt{\#ebola} as at least one 
of the hashtags.  The hashtags-vs-counts in a log-log scale plot 
is shown in Figure \ref{fig:hashtag_counts}.  We also provide a list of the top 20
hashtags along with their counts in Table \ref{tab:hashtag_counts}\footnote{The full
list of hashtags and their counts can be found here: 
\small{\url{https://github.com/kjanani/KDD\_BigCHat2015}.} under the \texttt{hashtags\_counts.txt} file.}.

We chose the top 1000 hashtags and manually sifted through them 
to form our set of \emph{credible} and \emph{speculative} hashtags with the 
intention that a tweet with a non-serious or speculative hashtag will 
likely have misinformation on Ebola, and a tweet with a credible 
hashtag will contain information that is accurate.  For each hashtag under 
consideration, we put them in one of three buckets: credible, speculative, and 
unsure.  Our criteria for a hashtag to be considered as credible were that
they should either indicate origin from a government agency such as the
Centers for Disease Control (e.g. \texttt{\#cdc}) or that they should indicate
origin from an authoritative source (e.g. \texttt{\#cnn}). 
We acknowledge that such hashtags could also be used convey that a 
certain information is not true.  Our argument here is that in 
such a situation, the tweet perhaps also has another hashtag
which is speculative in nature.
In that case, those tweets will be discarded from the data.

On the other hand, to be a good candidate for the 
speculative list, we looked for something in the hashtag that was 
indicative of one of the following characteristics: humor (e.g. \texttt{\#ebolazombie, \#twitterjoke, 
\#ebolajokes}), sarcasm (eg. \texttt{\#andnowihaveebola, \#gossip}), fear (\texttt{\#scarystoriesin5words}), 
or any indications of the tweet being a rumor (\texttt{\#lies, \#falsenews}).
If a hashtag did not definitively fall under either 
of the two buckets, it was placed in the unsure bucket, and essentially discarded.

Our list of \emph{credible} hashtags and \emph{speculative} hashtags are provided in 
Table \ref{tab:credible_speculative}.
As mentioned earlier, since a tweet can contain more than one 
hashtag, there is a possibility that a tweet may contain 
hashtags from both the \emph{credible} and the \emph{speculative} set.  
No such tweets were included in this study.  The \emph{credible} set 
contained approximately 89K tweets and
the \emph{speculative} set contained approximately 20K tweets.  

\subsection{Feature Extraction}
As a preliminary experiment, we extracted several Twitter based features
to the discover the differentiating factors between the \emph{credible} and the
\emph{speculative} set of tweets.  These features have interesting social
interpretations.  For example, the presence of a \texttt{url} could indicate
that the certain tweet links to a source validating the information provided
in the tweet \footnote{Some studies on spam detection have also found that
the tweet from a spammer is highly likely to have a link embedded in it \cite{Lee_crowdturfers}.}.
Similarly, a user with a large number of followers and a relatively lower number
of followees could indicate that he/she is a celebrity.  In addition to studying
the numerical differences between features of the \emph{credible} and the
\emph{speculative} set, our focus here is to obtain meaningful conclusions about 
the semantic and social characteristics of the two sets as well. 

We performed statistical significance tests to assess 
if the two sets indeed were generated from distributions with different means 
and variances using two-sample t-tests.  A sublist of the features\footnote{A complete list of all the
extracted features can be found here: \small{\url{https://github.com/kjanani/KDD\_BigCHat2015}} under
the \texttt{column\_names\_new.txt} file.},
their descriptions and the statistical significance test results are
provided in Table \ref{tab:ttests}.  

\section{Results and Discussion}
\begin{table*}
\centering
\begin{tabular} {|l|l|l|l|l|l|}
\hline
\textbf{feature} & \textbf{description} & \textbf{\emph{credible} avg.} & \textbf{{\emph{speculative}} avg.} & \textbf{hyp} & \textbf{p-value}\\
\hline
\texttt{retweeted\_status} & \small{[0/1] feature indicating if the tweet is a retweet}& 0.4378 & 0.5507 & 1 & 4.41E-176 \\
\texttt{retweet\_count} & \small{number of times the tweet has been retweeted}& 189.3917 & 78.3532 & 1 & 0 \\
\texttt{favorite\_count} & \small{number of times the tweet has been favorited}& 59.8401 & 33.1443 & 1 & 5.25E-212 \\
\texttt{in\_reply\_to\_status\_id} & \small{[0/1] feature indicating if the tweet is a reply}& 0.0727 & 0.0267 & 1 & 1.95E217 \\
\texttt{entities\_urls} & \small{number of urls in the tweet}& 0.4574 & 0.4226 & 1 & 2.39E-17 \\
\texttt{entities\_symbols} & \small{number of \$ symbols in the tweet}& 0.0114 & 0.0005 & 1 & 6.96E-25 \\
\texttt{entities\_hashtags} & \small{number of hashtags in the tweet}& 2.7971 & 2.3286 & 1 & 8.03E-297 \\
\texttt{user\_verified} & \small{[0/1] feature indicating if the user is verified} & 0.0176 & 0.0014 & 1 & 1.51-214 \\
\texttt{user\_friends\_count} & \small{number of followees for the user}& 1619.29 & 1741.45 & 0 & 0.0517\\
\texttt{user\_followers\_count} & \small{number of followers for the user}& 6960.54 & 2635.26 & 1 & 1.69E-28 \\
\texttt{user\_status\_count} & \small{number of status updates by the user}& 36241.94 & 19549.95 & 1 & 0 \\
\texttt{user\_favourites\_count} & \small{number of favourites received by the user}& 2620.60 & 2237.58 & 1 & 0 \\
\texttt{possibly\_sensitive} & \small{indicates if the media is sensitive}& 0.0134 & 0.0157 & 0 & 0.0157 \\
\hline
\end{tabular}
\caption{\textbf{\small{This table illustrates a sublist of features, their descriptions,
the average values in the \emph{credible} and the \emph{speculative} sets, and results
from statistical significance tests. A value of $\textbf{p} \leq 0.002$ is considered as the breakpoint.}}}
\label{tab:ttests}
\end{table*}

In this section, we draw some conclusions from the information
listed in Table \ref{tab:ttests}.

The average number of hashtags per tweet in the \emph{credible} set 
is 2.79 and in the \emph{speculative} set is 2.32 (Table \ref{tab:ttests}, \texttt{entities\_hashtags}).  
Note that both averages would have to be greater than one because of the 
way we chose the tweets for the two sets: we made sure that the tweet 
has at least one hashtag.  As it turns out, the number of hashtags per 
tweet for the \emph{credible} set is greater than the number of hashtags per 
tweet for the \emph{speculative} set.  This possibly indicates that the credible 
tweets have a better context in which they can be placed, and hence the use
of more hashtags. 

The average number of urls per tweet (Table \ref{tab:ttests}, \texttt{entities\_urls})
for the \emph{credible} set is also greater 
than the average number of urls per tweet for the \emph{speculative} set.  
This indicates that the credible tweets typically have some form of 
validation to support their argument, as opposed to the speculative tweets.  

There is also a greater percentage of verified users in the \emph{credible} set
(Table \ref{tab:ttests}, \texttt{user\_verified}).  
This perhaps simply indicates that the verified users seem to endorse 
tweets that seem more credible than otherwise.  

The average number of followers (\texttt{user\_followers\_count})
for the users in the \emph{credible} set is higher than for the users in the \emph{speculative} set.  
The number of followers for the average user in the \emph{speculative}
set is $\approx$ 2600, and the number of followers for the 
average user in the \emph{credible} set is $\approx$ 7000; approximately 2.6 times the former.  This stark difference 
between the two values could be a correlated effect of 
observing a greater percentage of verified users (\texttt{user\_verified}) in the \emph{credible} set.  
Verified users typically tend to have many more 
followers than the average user, and hence justifying
the presence of a greater number of followers for the average user in the \emph{credible} set.

The \texttt{retweeted\_status} is a [0/1] feature which indicates if the tweet is a retweet.
This feature is lower for the \emph{credible} set when compared to the
\emph{speculative} set which might indicate that the tweets in the credible set
more often contain new information.  Furthermore, when analyzing this feature in conjunction with
the results of the feature \texttt{retweet\_count} we observe that the difference is
much higher between the \emph{credible} and the \emph{speculative} set.  That is,
the total number of retweets in the \emph{credible} are fewer, 
but they are being retweeted several more times in the \emph{credible}
set than for the \emph{speculative} set.  This possibly suggests the presence of
fewer important messages in the \emph{credible} set, but they spread widely in the network.
However, there are several more speculative messages which do not spread very widely.

An interesting feature is the \texttt{possibly\_sensitive} feature.  
This is a binary feature indicating whether there is sensitive
material in the media uploaded by the user.  It is up to the user's
discretion to mark the media as sensitive.  Since this feature
is specific to the media uploaded by the user, this feature becomes irrelevant
in those tweets without a url (all uploaded media
appear in tweets via their url).  Although this feature does not exhibit as stark a difference as the other features
($p$ = 0.0157), it is interesting to note that the \emph{credible} set of tweets
has a lesser percentage of sensitive media than the \emph{speculative}
set.


\section{Contributions and Limitations}
The main contribution of this study is a 
proposed method to differentiate between \emph{credible} 
and \emph{speculative} tweets using hashtags as 
category indications. We have collected a dataset 
of tweets from the Twitter streaming API during a 
crucial time-period that is a valuable resource for 
studying (mis-)information spread to the public. 
We also show that structured features from the Twitter 
API can be useful for this task. One limitation in 
this study is the manual division of hashtags into these 
two sets; further investigation is needed to verify that 
this division reflects an accurate distinction between \emph{credible} 
and \emph{speculative} tweets. A further limitation is the fact 
that only tweets with hashtags are used; tweets without hashtags 
would need to be annotated in order to know which category they fall into.

\section{Future Directions}
The data analysis methodology 
for what could possibly be credible and mis-informative has shown 
promising preliminary results.

In the future, we plan to explore several other features.  Most
of the features considered in this study were network features.
It would be interesting to analyze the content of the tweets,
using Natural Language Processing techniques.  For example, although the probability
of actually contracting the virus was very minute, a huge majority of
the public feared contracting the virus.  It would be interesting
to see if this had any effects on the general sentiment of the
Twitter feeds.  

Another important goal of this study is to locate the source of false
rumors and counter their spread as early as possible.  Hence, a more
temporal analysis could help both in localizing the rumor sources
and countering them in the early stages. 
\bibliographystyle{plain}
\bibliography{refs}

\begin{thebibliography}{10}

\bibitem{time:2014}
Fear, misinformation, and social media complicate ebola fight.
\newblock \url{http:time.com/3479254/ebola-social-media}.

\bibitem{Tsur:2012}
What's in a hashtag?: content based prediction of the spread of ideas in
  microblogging communities.
\newblock In {\em WSDM}, pages 643--652. ACM, 2012.

\bibitem{anema2014digital}
Aranka Anema, Sheryl Kluberg, Kumanan Wilson, Robert~S Hogg, Kamran Khan,
  Simon~I Hay, Andrew~J Tatem, and John~S Brownstein.
\newblock Digital surveillance for enhanced detection and response to
  outbreaks.
\newblock {\em The Lancet Infectious Diseases}, 14(11):1035--1037, 2014.

\bibitem{Castillo:2011}
Carlos Castillo, Marcelo Mendoza, and Barbara Poblete.
\newblock Information credibility on twitter.
\newblock In {\em Proceedings of the 20th International Conference on World
  Wide Web}, WWW '11, pages 675--684, New York, NY, USA, 2011. ACM.

\bibitem{DeChoudhury:2014}
Munmun De~Choudhury, Scott Counts, Eric~J. Horvitz, and Aaron Hoff.
\newblock Characterizing and predicting postpartum depression from shared
  facebook data.
\newblock In {\em Proceedings of the 17th ACM Conference on Computer Supported
  Cooperative Work \&\#38; Social Computing}, CSCW '14, pages 626--638, New
  York, NY, USA, 2014. ACM.

\bibitem{doan2012enhancing}
Son Doan, Lucila Ohno-Machado, and Nigel Collier.
\newblock Enhancing twitter data analysis with simple semantic filtering:
  Example in tracking influenza-like illnesses.
\newblock In {\em Healthcare Informatics, Imaging and Systems Biology (HISB),
  2012 IEEE Second International Conference on}, pages 62--71. IEEE, 2012.

\bibitem{doan2012analysis}
Son Doan, Bao-Khanh~Ho Vo, and Nigel Collier.
\newblock An analysis of twitter messages in the 2011 tohoku earthquake.
\newblock In {\em Electronic Healthcare}, pages 58--66. Springer, 2012.

\bibitem{eysenbach2006infodemiology}
Gunther Eysenbach.
\newblock Infodemiology: tracking flu-related searches on the web for syndromic
  surveillance.
\newblock In {\em AMIA Annual Symposium Proceedings}, volume 2006, page 244.
  American Medical Informatics Association, 2006.

\bibitem{eysenbach2009infodemiology}
Gunther Eysenbach.
\newblock Infodemiology and infoveillance: framework for an emerging set of
  public health informatics methods to analyze search, communication and
  publication behavior on the internet.
\newblock {\em Journal of medical Internet research}, 11(1), 2009.

\bibitem{Gupta:WWW_2013}
Aditi Gupta, Hemank Lamba, Ponnurangam Kumaraguru, and Anupam Joshi.
\newblock Faking sandy: Characterizing and identifying fake images on twitter
  during hurricane sandy.
\newblock In {\em Proceedings of the 22Nd International Conference on World
  Wide Web Companion}, WWW '13 Companion, pages 729--736, Republic and Canton
  of Geneva, Switzerland, 2013. International World Wide Web Conferences
  Steering Committee.

\bibitem{hartley2010landscape}
DM~Hartley, NP~Nelson, Ronald Walters, Ray Arthur, Roman Yangarber, Larry
  Madoff, Jens Linge, Abla Mawudeku, Nigel Collier, John Brownstein, et~al.
\newblock The landscape of international event-based biosurveillance.
\newblock {\em Emerging Health Threats Journal}, 3(e3), 2010.

\bibitem{Lee_crowdturfers}
Kyumin Lee, Prithivi Tamilarasan, and James Caverlee.
\newblock Crowdturfers, campaigns, and social media: Tracking and revealing
  crowdsourced manipulation of social media.

\bibitem{myslin2013using}
Mark Mysl{\'\i}n, Shu-Hong Zhu, Wendy Chapman, and Mike Conway.
\newblock Using twitter to examine smoking behavior and perceptions of emerging
  tobacco products.
\newblock {\em Journal of medical Internet research}, 15(8), 2013.

\bibitem{oyeyemi2014ebola}
Sunday~Oluwafemi Oyeyemi, Elia Gabarron, and Rolf Wynn.
\newblock Ebola, twitter, and misinformation: a dangerous combination?
\newblock {\em BMJ}, 349:g6178, 2014.

\bibitem{rolison2015knowledge}
Jonathan~J Rolison and Yaniv Hanoch.
\newblock Knowledge and risk perceptions of the ebola virus in the united
  states.
\newblock {\em Preventive Medicine Reports}, 2:262--264, 2015.

\bibitem{xie2014correlation}
Tiansheng Xie, Zongxing Yang, Shigui Yang, Nanping Wu, and Lanjuan Li.
\newblock Correlation between reported human infection with avian influenza a
  h7n9 virus and cyber user awareness: what can we learn from digital
  epidemiology?
\newblock {\em International Journal of Infectious Diseases}, 22:1--3, 2014.

\end{thebibliography}
\end{document}